\documentclass[titlepage,12pt]{article}
\usepackage{myarticle}

\newfam\mbffam \def\mbf#1{{\mathbf#1}}
\def\avg#1{\langle #1 \rangle} 
\def\tr{\mathrm{tr}} 
  \def\Vx{\mbf x}

\font\tenscr=rsfs10

\font\twelvescr=rsfs10 scaled \magstep 1
\skewchar\tenscr='177
\newfam\scrfam \textfont\scrfam=\twelvescr
\def\scr#1{{\fam\scrfam\relax#1}}
\def\SV{{\scr V}} 

\begin{document}
\preprint{UTTG-15-98}
\title{\Large\bf  New skyrmions in the attractive Hubbard model 
with broken SO($4$) symmetry}

\author{W. Vincent Liu\footnote{Email: liu@physics.utexas.edu} } 
\address{Theory Group, Department of Physics, University of Texas,
Austin, Texas 78712} 
\Abstract{The coexistence of superconducting and
charge-density-wave order in the half-filled attractive Hubbard model
is interpreted as a consequence of the pseudospin SU($2$) symmetry
spontaneously broken to a `hidden'
subgroup U($1$).  
By topological arguments we show that there must exist
new skyrmion textures associated with this symmetry breakdown.
This fact is illustrated via a non-linear $\sigma$-model.
Unlike the spin textures previously known
in an antiferromagnetic background,
doping the model away from half-filling leads
the new skyrmions to unwrap.
\vspace{5pt}
}
\maketitle
 
Recently, the Hubbard model
has been a subject of intense study for its possible applications to 
high $T_c$ superconductivity \cite{Anderson:87}.
It represents the simplest theoretical framework for describing
strongly correlated electron systems. 
Despite its simple appearance, it is however quite
difficult to analyze beyond one dimension. 
The elimination of phonon or other internal degrees of freedom can
give rise to an effective attraction between the remaining electrons
\cite{Anderson:75}.
In this paper, we shall restrict ourselves to the 
Hubbard model of attractive interaction, 
which has been frequently applied to 
such compounds  as bismuth oxide superconductors
\cite{ref:bismuth} and many others \cite{Czart+:96}.    
Numerical studies \cite{Scalettar+:89} showed 
that for dimension $d\geq 2$ it has a  ground state of both
charge-density wave (CDW) order and superconductivity (SC) 
when half-filled.   
We shall show how an effective action for low energy excitations 
around such a ground state can be derived directly from
the spontaneous breakdown of the SO($4$)
symmetry of the model
\cite{Affleck:89,Yang-Zhang:90,Schulz:90}, and that there must 
exist a new kind of skyrmion textures associated with this breakdown.

Let us consider the $d$-dimensional ($d=2$ or $3$)
Hubbard model on a square or cubic lattice. The 
Lagrangian  reads (summation convention on repeated indices assumed) 
\begin{equation}
L= \sum_{\Vx} c^\dag_\sigma(\Vx,t) i\partial_t
c_\sigma(\Vx,t)  + t\sum_{\avg{\Vx\Vx^\prime}} \left[
c^\dag_{\sigma}(\Vx,t)c_{\sigma}(\Vx^\prime,t)  + h.c. \right] - U\sum_{\Vx}
n_\uparrow(\Vx,t)n_\downarrow(\Vx,t) \quad ,  \label{eq:L}
\end{equation}
where $c_\sigma(\Vx,t)$ and $c^\dag_\sigma(\Vx,t)$ respectively 
annihilates and
creates electrons with spin $\sigma$ ($=\uparrow,\downarrow$) at
site $\Vx=(x_1,\cdots,x_d)$ and time $t$, and
$n_\sigma(\Vx)=c^\dag_{\sigma}(\Vx)c_{\sigma}(\Vx)$ (no summation). 
Hopping is restricted to nearest neighbors, as indicated by the bracket
$\avg{\Vx\Vx^\prime}$, with $t$ the (constant) 
transferring matrix element. The constant $U$ is the on-site Hubbard
interaction. This is
a standard theory for electrons in well-localized atomic orbitals with a
probability $t$ for transitions between neighboring atoms. 
By now it is well known that the Hubbard model 
at half-filling has SO($4$)$\simeq$
SU(2)$_C\times$ SU(2)$_S/Z_2$ symmetry
\cite{Affleck:89,Yang-Zhang:90,Schulz:90}, 
where SU(2)$_S$ corresponds to
spin rotational invariance and SU(2)$_C$ (sometimes called
pseudospin group \cite{SCZhang:90}) 
contains the ordinary U$(1)_C$ charge symmetry as a
subgroup.  
To see the SO($4$) symmetry of the Lagrangian (\ref{eq:L}) 
in a manifest way, we
use the matrix representation as introduced by Affleck
\cite{Affleck:89} and Schulz \cite{Schulz:90},
\begin{equation}
\Psi(\Vx,t) =\left(
\begin{array}{cc}
c_\uparrow(\Vx,t) & c_\downarrow(\Vx,t)\\
(-)^\Vx c^\dag_\downarrow(\Vx,t) & -(-)^\Vx c^\dag_\uparrow(\Vx,t)
\end{array} \right) \quad , \label{eq:Psi}
\end{equation}
where we have adopted the notation $(-)^\Vx\equiv (-)^{\sum_i x_i}$.  (As
usual, the lattice spacing  is taken to be unit.) 
An ordinary spin SU$(2)_S$ transformation corresponds to right
multiplication of $\Psi$ by an SU$(2)_S$ matrix $U$: 
$\Psi \rightarrow \Psi U$, while the pseudospin SU$(2)_C$
transformation 
is left multiplication: $\Psi \rightarrow \tilde{U}\Psi $. Obviously, 
the latter generates the electron-hole transformation.
At half-filling, defined by $\sum_{\Vx\sigma} n_\sigma(\Vx) = 
\sum_{\Vx} 1$, the Lagrangian can  be rewritten in terms of $\Psi$:
\begin{eqnarray}
L &=& \sum_{\Vx} {1\over 2}
\tr[\Psi^\dag i\partial_t\Psi] 
+t \sum_{\avg{\Vx\Vx^\prime}} \tr [\Psi^\dag(\Vx,t)\Psi(\Vx^\prime,t)]
\nonumber \\ 
&&
- {U\over 24} \sum_{\Vx a} \left(\tr[(-)^\Vx\Psi^\dag \sigma_a  
\Psi]\right)^2 
\quad , \label{eq:L2}
\end{eqnarray}
where $\sigma_{a}$ ($a=1,2,3$)  are the Pauli matrices.
The theory now is manifestly SU(2)$_C\times$ SU(2)$_S$ invariant.
The interaction term in Eq.~(\ref{eq:L2}) 
is devised in such to our later purpose of studying superconductivity
and CDW order (although obviously it can be
written in several equivalent forms).
By Noether's method we may derive from this Lagrangian the conserved
spin and pseudospin charges, 
\begin{eqnarray}
\vec{S} &=& \sum_\Vx {1\over 4}\tr[\Psi^\dag \Psi\vec{\sigma}] 
=\sum_\Vx{1\over 2}\{c^\dag_\uparrow c_\downarrow +h.c. \, , 
		 ic^\dag_\uparrow c_\downarrow +h.c. \, ,
		c^\dag_\uparrow c_\uparrow -c^\dag_\downarrow
c_\downarrow \} ,   \label{eq:spinS}
\\
\vec{J}&=& \sum_\Vx {1\over 4}\tr[\Psi^\dag\vec{\sigma} \Psi] 
=  
\sum_\Vx{1\over 2}\{(-)^\Vx(c_\downarrow c_\uparrow +h.c.) \, , 
		 (-)^\Vx(ic_\downarrow c_\uparrow +h.c.) \, ,
		c^\dag_\sigma c_\sigma -1 \}, \label{eq:pseudospinJ}
\end{eqnarray} 
which are the
generators respectively of spin and pseudospin symmetries. 
($J_3$ may be identified as the
ordinary U$(1)_C$ charge generator apart from a constant.) 
The quantum operators $\vec{S}$ and $\vec{J}$ satisfy the commutation
relations: 
$[{S}_a,{S}_b] =i \epsilon_{abc} {S}_c$, 
$[{J}_a,{J}_b] =i \epsilon_{abc} {J}_c$, and
$[S_a,{J}_b]=0$.

The SO($4$) symmetry can be spontaneously broken in various patterns
(see Table~\ref{tab:comparison}).
Several authors \cite{Shraiman-Siggia:88,Gooding:91,Seibold:98pre}
showed that  spin textures (skyrmions)
can exist in an antiferromagnetic
(N\'eel) background on a square lattice.   
In fact, such skyrmions are 
associated with the spontaneously breakdown of the spin rotational
symmetry SU$(2)_S$  to U$(1)_S$, as exhibited by the
O($3$) $\sigma$-model \cite{remark:Fradkin:bk91}.    Yet, 
we shall explore an alternative case of SU$(2)_C$ broken to
a `hidden' U($1$) subgroup (not U$(1)_C$ in general), for which  the
superconducting  and CDW long-range orders coexist
\cite{Scalettar+:89}.  

We now proceed to derive an effective field theory of low energy
excitations. In
the path-integral formalism, the action functional is
given by the same form as Eq.~(\ref{eq:L2}) except replacing the
electron annihilation and creation operators by the Grassmann 
fields (acting like 
anticommuting $c$-numbers): $c_\sigma \rightarrow \psi_\sigma$ and 
$c^\dag_\sigma \rightarrow \psi^\dag_\sigma$. 
By Hubbard-Stratonovich transformation we can cancel the four-fermion
interaction term in Eq.~(\ref{eq:L2}) by adding to the Lagrangian a term
$\Delta L={-2|U|\over 3}\sum_{\Vx a}(\phi_a-{1\over 4} 
\tr[(-)^\Vx\Psi^\dag \sigma_a \Psi])^2$ (for negative $U$), 
where we have introduced  
three auxiliary scalar fields $\phi_a$. Then,
\begin{eqnarray}   
L &=&\sum_{\Vx} \left\{
\psi^\dag_\sigma i\partial_t\psi_\sigma + {2U\over 3}
{\phi}_a\phi_a 
- {U\over 3}\phi_a \tr[(-)^\Vx\Psi^\dag\sigma_a  
\Psi] \right\}
 \nonumber \\ 
&&+t \sum_{\avg{\Vx\Vx^\prime}}
\left[\psi^\dag_\sigma(\Vx,t)\psi_\sigma(\Vx^\prime,t) +h.c.\right]  
\quad . \label{eq:L3}
\end{eqnarray}
Integrating out the $\phi_a$ fields shall return $L$ to the original
form. 
The Lagrangian (\ref{eq:L3}) is stationary in `field' space at the
point
\begin{eqnarray}
\vec{\phi}_{0}(\Vx,t)& =&{1\over 4}\tr [(-)^\Vx\Psi^\dag \vec{\sigma} \Psi] 
\nonumber \\ 
 &=& {1\over 2} \left\{ \psi_\downarrow \psi_\uparrow+h.c., \,
		i\psi_\downarrow\psi_\uparrow +h.c., \,
		(-)^\Vx \psi^\dag_\sigma\psi_\sigma
 \right\}  . \label{eq:phi0}
\end{eqnarray}
The expression of $\vec{\phi}_0$ infers that $\vec{\phi}$ is a
three-vector in  pseudospin
space but  transforms as a scalar under the
spin group SU$(2)_S$ \cite{remark:phi}. 

Now that the Lagrangian (\ref{eq:L3}) 
is quadratic in fermion fields $\psi_\sigma$ and
$\psi^\dag_\sigma$, one can formally integrate out all of them and
obtain the effective action  for the fields of $\phi_a$, which 
represent the collective modes
associated with `pseudospin' fluctuations.
Along this direction, one can follow a standard procedure 
(see, e.g., Sec.~3.4 of
Ref.~\cite{remark:Fradkin:bk91}).
However,  we shall derive
the low energy effective field theory with an alternative approach, in
which the role of symmetry breaking will be evident, based on
such an argument: the Lagrangian 
(\ref{eq:L3}) is  SU$(2)_C$
invariant, and hence any effective action derived from it must be also
SU$(2)_C$ invariant (no matter whether realized linearly or nonlinearly). 
Thus, together with time-reversal symmetry and parity, the SU$(2)_C$
invariance requires that
the (continuum-limit) effective action for $\vec{\phi}$ 
be given in the following general form
\begin{eqnarray}
I_{{\mathrm{eff}}}&= &\int d^dx dt \left[ -c_1\vec{\phi}\cdot
i\partial_t\vec{\phi} + {1\over 2} c_2 \partial_t\vec{\phi} \cdot
\partial_t\vec{\phi} - {1\over 2} c^\prime_2 \partial_i\vec{\phi} \cdot
\partial_i\vec{\phi}  + \cdots -\SV(\vec{\phi}) \right], \label{eq:Ieff}  \\
\SV(\vec{\phi}) &=&  d_2 \vec{\phi}\cdot\vec{\phi} + 
d_4 (\vec{\phi}\cdot\vec{\phi})^2 +\cdots , \label{eq:V}
\end{eqnarray}
where space index $i$ runs from $1$ to $d$.
The terms indicated by `$\cdots$' will contain higher powers of 
the  $\vec{\phi}$ fields and/or their derivatives. All coefficients ($c_1$,
$\cdots$, and $d_2$, $\cdots$) are  assumed real and finite after
renormalization \cite{remark:eft}.

With the ground state pointing in some specific direction in
pseudospin space $\avg{\vec{\phi}}=\vec{\Delta}\neq 0$,
the pseudospin SU$(2)_C$ symmetry is spontaneously broken down,
whereas the ordinary spin SU$(2)_S$ rotational invariance is
preserved but rather realized trivially. (Recall
that $\vec{\phi}$ and $\vec{\Delta}$ transform as scalars under SU$(2)_S$.)
By Eq.~(\ref{eq:phi0}), 
\begin{equation}
\Delta_1\equiv {1\over 2}\avg{\psi_\downarrow\psi_\uparrow 
+ \psi^\dag_\uparrow\psi^\dag_\downarrow},\quad
 \Delta_2\equiv {i\over 2}\avg{\psi_\downarrow\psi_\uparrow 
 -\psi^\dag_\uparrow\psi^\dag_\downarrow}\quad \mbox{and}\quad
\Delta_3\equiv {1\over 2} \avg{(-)^\Vx\psi^\dag_\sigma\psi_\sigma},
\label{eq:Delta}
\end{equation}
and they are naturally interpreted as the 
pairing fields and CDW order parameter. This is a manifestation of the
fact that CDW order and superconductivity can coexist in the
negative-$U$ Hubbard model at half-filling
for $d\geq 2$ \cite{Scalettar+:89}.
Nevertheless, the ground state of $\avg{\vec{\phi}}=\vec{\Delta}$ 
is still invariant under  SO($2$) $\simeq$ U($1$) rotations about the
direction of $\vec{\Delta}$. This unbroken `hidden' U($1$) subgroup
is however generally not the same as the ordinary U$(1)_C$ charge
symmetry.   
Clearly, a transformation of SU$(2)_C$ can rotate the SC pairing fields
$\Delta_{1,2}$ and the CDW order parameter $\Delta_3$ into one another
\cite{Wallington-Annett:97}. 
Thus, the coexistence of SC and CDW can be perfectly understood as a
result of the breakdown of SU$(2)_C$ $\rightarrow$
U$(1)_{\mathrm{hidden}}$ with the order-parameter space 
topologically regarded as  a $2$-sphere manifold $S_2$.  

Since the action (\ref{eq:Ieff}) is SU$(2)_C$ invariant, it is always
possible to rotate
the pseudospin  `field' space in such a way that $\vec{\Delta}$ aligns
into the three-direction ($z$-direction) 
in the new `field' space.
We then express the $\vec{\phi}$ field as a
pseudospin rotation $R$ acting on a three-vector $(0,0,\sigma)$ whose
first two components vanish: $\phi_a(\Vx,t) = R_{a3}(\Vx,t)
\sigma(\Vx,t)$ with 
$R_{ab}$ ($a,b=1,2,3$) an orthogonal matrix $R^T(\Vx,t) R(\Vx,t) =1$. 
In place of the field variables $\phi_a$, our  variables  now  are
$\sigma(\Vx,t)$ and whatever other variables are needed to
parameterize the rotation $R$. Let us simply choose those  parameters  
as the $R_{a3}(\Vx,t)$ themselves but rather 
call them  $\Omega_a(\Vx,t)$ hereafter. 
By definition, $\Omega_a$ represent the Goldstone mode degrees of
freedom for the pseudospin fluctuations. 
Furthermore,  as far as the lowest energy excitations
are concerned, we may simply replace $\sigma(\Vx,t)$ with its
mean-field value $|\vec{\Delta}|\equiv \sqrt{\Delta_a\Delta_a}$. 
The effective action (\ref{eq:Ieff}) then becomes
\begin{equation}
I_{\mathrm{eff}} =\int d^dx dt  {c_2|\vec{\Delta}|^2 \over
2} \left[ (\partial_t\vec{\Omega})^2 - {c^\prime_2\over c_2}
\partial_i\vec{\Omega}\cdot\partial_i\vec{\Omega}
  + \cdots \right] \label{eq:Ieff2}
\end{equation}
with a constraint $\vec{\Omega}\cdot\vec{\Omega}=1$. 
This is the familiar nonlinear O($3$) $\sigma$-model. It
exhibits that two gapless Goldstone modes appear around
the CDW and superconductivity background along with the spontaneous
breakdown of  SU$(2)_C\times$SU$(2)_S$ $\rightarrow$
U$(1)\times$SU$(2)_S$, as required by the
Goldstone theorem \cite{Goldstone:Weinberg:bk96:ch19}.  
We remark that the result is true for  
$d\geq 2$ space dimensions only
\cite{remark:Mermin-Wagner-Coleman}. 

The effective field theory described by Eq.~(\ref{eq:Ieff2}) 
can predict the
existence of a topological object, known as
skyrmion. To see this, consider the energy of a static solution
\begin{equation}
E= {c^\prime_2|\vec{\Delta}|^2\over 2} \int d^dx \partial_i\Omega_a \cdot
\partial_i{\Omega_a} +\cdots . \label{eq:E}
\end{equation}
Field configurations of finite $E$ must have
$\partial_i\Omega_a(\Vx)$ vanishing at spatial infinity faster than
$|\Vx|^{-d/2}$ (where $|\Vx|\equiv \sqrt{x_ix_i}$), so that
$\Omega_a(\Vx)$ must approach  a constant $\Omega_{a\infty}$ as
$\Vx\rightarrow \infty$ with a remainder vanishing faster than
$|\Vx|^{(2-d)/2}$. The fields $\Omega_a$ at any point form a
homogeneous space, the (order-parameter) coset space of
$\mathrm{SU}(2)_C\times \mathrm{SU}(2)_S / \mathrm{U}(1)\times
\mathrm{SU}(2)_S \simeq \mathrm{SO}(3)/\mathrm{SO}(2) =
 S_2$ ($2$-sphere), for which it is possible to
transform any one field value to any other by a transformation of
SU$(2)_C$. The field $\Omega_a(\Vx)$ thus represents a mapping of the
whole $d$-dimensional space, with the sphere $r=\infty$ taken as a
single point, into  the manifold $S_2$ of all field values. 
Therefore, finite-energy static configurations $\Omega_a(\Vx)$ in $d$
space dimensions  may be classified according to the $d$th homotopy
group of $S_2$, $\pi_d(S_2)$ (for a lucid discussion, see, e.g.,
Chap. 23 of Ref.~\cite{Goldstone:Weinberg:bk96:ch19}). 

First, let us consider $d=2$ case.  The second homotopy group
$\pi_2(S_2)=Z$ where $Z$ is the group of integers. Therefore, there
exist skyrmions characterized by an integral winding number $Q$
(equally called topological charge as well), which can be expressed as
$$
Q={1\over 8\pi} \int d^2x \epsilon_{ij}\vec{\Omega} \cdot
(\partial_i\vec{\Omega} \times \partial_j\vec{\Omega}).
$$ 
A skyrmion
with winding number $Q$ has energy $E=4\pi c^\prime_2|\vec{\Delta}|^2
|Q|$, and its analytical solution is already known 
\cite{remark:Rajaraman:82}. 
For example, the solution for a skyrmion of winding number 
$Q=1$ and  scale $\lambda$ reads 
$$
\Omega_{1,2}(\Vx)={4\lambda x_{1,2}\over \Vx^2+ 4\lambda^2}, \quad
\mbox{and} \quad
\Omega_{3}(\Vx)={\Vx^2-4\lambda^2 \over \Vx^2+ 4\lambda^2}.
$$
As of $d=3$, $\pi_3(S_2)=Z$. Hence, the
theory also predicts the existence of pseudospin textures (skyrmions)
classified by an integral winding number. Strictly speaking, the
functional (\ref{eq:E}) however does not have skyrmion stationary
points for $d=3$, unless we add terms involving higher powers of
$\partial_i\Omega_a$ to the integrand, as indicated by `$\cdots$'. For
instance, a four-derivative term will be sufficient \cite{remark:Derrick}.

One can gain an insight into the pseudospin skyrmions by comparing the
pseudospin $\vec{J}$ with the ordinary spin $\vec{S}$ as defined in
Eqs.~(\ref{eq:pseudospinJ}) and (\ref{eq:spinS}), respectively. 
The Hubbard Lagrangian
(\ref{eq:L}) is invariant under the combined action of the
Lieb--Mattis canonical transformation defined by
$c_\uparrow(\Vx) \rightarrow  c_\uparrow(\Vx)$, 
$c_\downarrow(\Vx) \rightarrow  (-)^{\Vx} c_\downarrow^\dag(\Vx)$, 
and the reversal of interaction sign
$U\rightarrow -U\,$\cite{Wallington-Annett:98}.
Meanwhile, the Lieb--Mattis  transformation maps
$\vec{J}$ onto $\vec{S}$. For $U>0$, the ground state is
antiferromagnetic with $\avg{\vec{S}}\neq 0$, which  spontaneously
breaks the ordinary spin symmetries, and the (spin)
skyrmions are known to exist. 
By contrast, when $U<0$, the ground state is in the phase of
SC and CDW with $\avg{\vec{J}}\neq 0$, which  spontaneously
breaks the pseudospin symmetries, and consequently there appear
pseudospin skyrmions.   (Table~\ref{tab:comparison} summaries the
analogy between the two cases.) 

The skyrmion configuration shall become unstable if the SO($4$)
symmetry is approximate. This happens when 
the Hubbard model is doped slightly away from
half-filling. Then,
a chemical potential term needs to be
included in the Lagrangian (Eqs.~\ref{eq:L} and \ref{eq:L2}): 
\begin{equation}
L_{\mathrm{break}} = -\mu\sum_\Vx c^\dag_\sigma c_\sigma 
 = - 2\mu J_3 + \mbox{Const.}\nonumber \,,
\end{equation}
which when $\mu\neq 0$ 
explicitly breaks the
SO($4$) symmetry down to the obvious U$(1)_C\times$SU$(2)_S$ subgroup.
Since the Lieb--Mattis  transformation maps
$J_3$ into $S_3$, this explicit breakdown of symmetry 
can be better viewed as 
caused by a `fictitious magnetic field',
namely, $\mu$. 
The presence of $L_{\mathrm{break}}$ 
will somehow  modify the effective action (\ref{eq:Ieff2})
and a detailed discussion shall be given elsewhere.
Nevertheless, according to our understanding of spontaneously broken
approximate symmetries, there are two immediate consequences of it. 
One is that the 
Goldstone modes become massive. (The usual phase mode
associated with U$(1)_C$ however remains massless.)
Another is that skyrmion configurations are allowed to unwrap. 
Imagine to do the following experiment. One
first creates skyrmions with some winding number
in the ground state and then
starts doping the system. We conjecture
that the `fictitious magnetic field' $\mu$ should cause the skyrmions
to unwrap but there will be some energy barrier to make it slow 
\cite{unwrapping:Weinberg}.  
By contrast, the spin skyrmion texture  in an
antiferromagnetic background  differs 
from the pseudospin skyrmion
here in that the former is topologically stable against doping.

I am indebted to my advisor Professor Steven Weinberg who has
been encouraging me to pursue this subject and sharing his insights
with me. Also, I thank J. Distler for patiently
teaching me homotopy theory  and T.-L. Ho for stimulating
conversations. 
Part of this work was done while visiting the Ohio
State University and I thank E. Braaten for the hospitality extended
to me.  This work is supported in part by NSF grant
PHY-9511632 and the Robert A. Welch Foundation.

\bibliographystyle{prsty} 
\bibliography{book,paper,skyrmion}

\begin{thebibliography}{10}
\bibitem{remark:phi}
By writing back $\phi_{0a}$ in terms of $c_\sigma$ and $c^\dag_\sigma$, one can
  easily verify that the resultant $\phi_{0a}$ operators form an irreducible
  tensor of rank $1$ under the SU$(2)_C$ algebra defined by the quantum
  operators $\vec{J}$ --- a result proved by Zhang \cite{SCZhang:90}.
\bibitem{remark:eft}
This is a standard argument in the effective field theory approach to
  spontaneous symmetry breaking. See, for instance, Weinberg
  \cite{Goldstone:Weinberg:bk96:ch19}.
\bibitem{remark:Mermin-Wagner-Coleman}
Recall that the Mermin--Wagner--Coleman theorem forbids the spontaneous
  breakdown of any continuous global symmetry in $1+1$ space-time dimensions at
  zero temperature.
\end{thebibliography}
\newpage
\begin{table}
\caption{Comparison of two possible patterns of symmetry
breaking. Here, the N\'eel vector
$\vec{M}\equiv\avg{(-)^\Vx\vec{S}(\Vx)}$ and $\vec{\Delta}$ is defined
in Eq.~(\ref{eq:Delta}).} 
\begin{center}
\begin{tabular}{lcc}
\hline
\multicolumn{1}{c}{Properties}  & $U>0$ & $U<0$  \\ \hline
Unbroken subgroup & SU$(2)_C\times$U$(1)_S$ &
U$(1)\times$SU$(2)_S$ \\
Broken symmetries & Two of $S_a$'s & Two of $J_a$'s \\
Phase & N\'eel & SC+CDW  \\
Order parameter &  $|\vec{M}|$  
& $|\vec{\Delta}|$\\    
Skyrmion ($d=2$) &  Spin texture & Pseudospin texture\\
Stability vs doping & Stable  & Unwrap \\
\hline
\end{tabular}
\end{center}
\label{tab:comparison}
\end{table}
\end{document}